\documentclass[sigconf]{acmart}

\AtBeginDocument{%
  }

\usepackage[utf8]{inputenc}
\usepackage{tabularx}
\usepackage{booktabs}
\usepackage{soul}
\usepackage{xcolor}
\usepackage{xspace}
\usepackage{graphicx}
\usepackage[caption=false]{subfig}  
\usepackage[export]{adjustbox}
\usepackage{amsmath}
\usepackage{textcomp}
\usepackage{listings}
\usepackage{cleveref}
\usepackage{algorithm}
\usepackage[noend]{algpseudocode}
\usepackage{balance}
\usepackage{enumitem}
\usepackage{multirow}
\usepackage{colortbl}
\usepackage{balance}
\usepackage{threeparttable}
\usepackage{multicol}
\usepackage{tcolorbox}

\definecolor{codegreen}{rgb}{0,0.6,0}
\definecolor{codegray}{rgb}{0.5,0.5,0.5}
\definecolor{codepurple}{rgb}{0.58,0,0.82}
\definecolor{backcolour}{rgb}{0.95,0.95,0.92}
\definecolor{gray}{gray}{0.9}
\definecolor{APA_stats}{RGB}{100, 100, 120}

\newcounter{observation}

\begin{document}

\title{Hidden Gems in the Rough: Computational Notebooks as an Uncharted Oasis for IDEs}

\author{Sergey Titov}
\email{sergey.titov@jetbrains.com}
\affiliation{%
  \institution{JetBrains Research}
  \city{Paphos}
  \country{Cyprus}
}

\author{Konstantin Grotov}\email{konstantin.grotov@jetbrains.com}
\affiliation{%
  \institution{JetBrains Research}
  \city{Munich}
  \country{Germany}
}

\author{Ashwin Prasad S. Venkatesh}
\email{ashwin.prasad@upb.de}
\affiliation{%
  \institution{Paderborn University}
  \city{Paderborn}
  \country{Germany}
}


\begin{abstract}

In this paper, we outline potential ways for the further development of computational notebooks in Integrated Development Environments (IDEs). We discuss notebooks integration with IDEs, focusing on three main areas: facilitating experimentation, adding collaborative features, and improving code comprehension. We propose that better support of notebooks will not only benefit the notebooks, but also enhance IDEs by supporting new development processes native to notebooks. In conclusion, we suggest that adapting IDEs for more experimentation-oriented notebook processes will prepare them for the future of AI-powered programming.

\end{abstract}


\newcommand{\ct}{~\hl{[]}\xspace}
\maketitle

\section{Computational Notebooks and IDE}

In recent years, computational notebooks have become a popular environment for writing code. They yield extensive application in numerous fields including, but not limited to, data science and analytics~\cite{perkel2018jupyter}, education~\cite{barba2019teaching}, science~\cite{perkel2021ten}, and much more.

The most popular computational notebooks platform~\cite{ecosystem2022} is an open-source project Jupyter~\cite{Kluyver2016jupyter}. 
Technically, Jupyter Notebook can be assumed to be an Integrated Development Environment (IDE) because it supports many programming languages, from popular ones like Python or C++ to new ones like Mojo~\cite{mojo}. It also offers various extensions to help improve the development process, for example, code completion extension or AI-assistant.\footnote{https://www.tabnine.com/install/jupyternotebook}
Moreover, Jupyter notebooks integrate data visualization tools and provide an interactive platform, making them highly practical for data analysis.

While the Jupyter Notebook is an IDE that can be improved accordingly, there is another direction of computational notebook evolution in the form of online services.
Several solutions offer a fresh perspective on enhancing the notebook workflow. These are not merely feature additions for coding but significantly unique approaches. For the comparison we chose Hex,\footnote{https://hex.tech} Datalore,\footnote{http://datalore.jetbrains.com} and DeepNote.\footnote{https://deepnote.com} All the mentioned services focused on improving collaborative aspects of working in notebooks, for example, by adding team chats, version control, and organization spaces. Also, they improve the developer's workflow by adding no-code widgets for working with data, for example, for convenient analysis of the data tables or data visualization purposes. Overall, these features allow users to organize experimental workflow within the team faster and spend less time on monotonous work with data. 

Although all the these solutions offer a convenient cloud-based platform for experimentation and collaboration, they are all similar to business intelligence systems (for example Redash\footnote{https://redash.io/} or Tableau\footnote{https://www.tableau.com/}) with the additional possibility of code development. As a result, these approaches come with a restricted range of options when choosing the environment --- \textit{e.g.}, users are limited to using specific versions of Python and have a finite number of built-in utilities. 

The online cloud-based approach showed its usefulness in the form of existing products, but it has its limitations. At the same time, we have modern, powerful IDEs, that solve these problems, providing rich customization and a much bigger amount of code features. 

In modern IDEs like Pycham or VScode, notebooks are supported. However, current integration is focused on enriching notebooks with IDE features like code completion or debugging tools without supporting the notebooks-specific workflows.

The unique and intertwined nature of development within notebooks frequently causes users to face challenges in their workflow, thereby disrupting the efficiency of their development processes~\cite{chattopadhyay2020s, ramasamy2023workflow}.
This situation underscores the need for the customization of integrated tools to address these specific challenges.
Consequently, Jupyter and notebooks, in general, should not be viewed merely as another form of a code editor in an IDE. Instead, they should be seen as a platform for conducting experiments, brainstorming, and exploration. This perspective suggests that notebooks demand tools not only to enhance the code but also to improve the overall development process, with a particular focus on experimentation and exploration.

We believe that the inclusion of the notebooks in IDE significantly increases the design space for both notebooks and IDEs, providing new possible features for both instruments.

\section{Exploring Notebooks in IDE Design Space}

The integration of the notebooks as a form of interaction within IDEs offers multiple opportunities for novel features that can rectify some of the inherent problems faced by web-based notebooks while simultaneously creating new and valuable functionalities for IDE users. This work concentrates on three key areas: enhancing the ease of experimentation, offering collaborative features, and bolstering the comprehensibility of notebooks.

\textbf{Facilitation of experimentation}. Notebooks are convenient tools for brainstorming and experimentation~\cite{rule2018exploration}.
A potential solution to enhance the experience could be streamlining the initiation of the experimentation process.
While Jupyter requires launching a server before starting work, and most IDEs necessitate setting up a project before granting access to the editor itself, we propose that streamlining the commencement of experimentation could significantly improve the user experience. Initiating a notebook should be as straightforward as opening a note or terminal.

A potential obstacle to a quicker start-up could be the management of environments. Rather than creating and configuring virtual environments separately, the IDE could offer a default environment associated with each programming language kernel for the notebook. This default environment would serve as the starting point for the user if no specific environment were chosen before beginning work in the notebook. This approach is commonly practiced in R notebooks, where the user operates with one primary environment and configures a new one only if necessary.\footnote{https://rmarkdown.rstudio.com/lesson-10.html} IDEs could take it one step further by providing users with a preconfigured environment tailored for different tasks, such as data analysis or machine learning.

Contrary to traditional approaches that add numerous functionalities, we believe in removing excess features and focusing on the essentials. We aim to streamline the notebook experience by minimizing environment setup time, allowing us to conduct experiments faster. By simplifying the interface and reducing unnecessary complexity, we seek to enhance user productivity and efficiency within notebooks. Without thinking about setting up the environment and the quality of the code, you can fully concentrate on the task and quickly solve it in a non-linear manner.

\textbf{Collaboration features}. The non-linearity of notebooks~\cite{ramasamy2023workflow} and the difficulties in their versioning~\cite{wang2019data} are significant challenges for convenient collaboration within a notebook. Online notebook platforms like Hex and Deepnote facilitate easier collaboration by allowing the creation of comment dialogues for selected cells or code elements. While a similar outcome can be achieved in offline notebooks through commentary in Markdown cells or inline code comments followed by sharing the notebook with commentary through a version control system, this approach is far from ideal. This type of commentary often lacks several important features.: 
\begin{itemize}
    \item Authorship -- while it still can be accessed by looking at who committed each comment, it still is very problematic to define, especially in a remote development setup
    \item Granularity -- markdown cells and comments, don't allow precisely pinpointing the target of the comment. In the notebooks, people may discuss specific parameters or resulting numbers.
    \item Dialog resolution and history -- while any text in the notebook could be saved using version control systems, there's no integrated way of tracking comment edits, resolutions, or mention of other collaborators. Collaborating and commenting in the notebooks should be as easy as in documents.
\end{itemize}

Developing a dedicated field in the cell metadata to store commentary dialogues and support their visualization in IDEs could solve these problems and enhance notebook communication. The most challenging aspect of this feature is establishing a unified standard for these dialogues to ensure compatibility across different computational notebook platforms—this requires collaborative efforts from the entire notebook community.

\textbf{Code Comprehension Enhancements}. 
Recent studies have found that notebooks tend to be longer and more entangled than scripts~\cite{grotov2022large}, making navigability and comprehension issues particularly critical for notebooks. To address these problems within an IDE, integrating automated documentation and structural text generation features could be a promising approach, as evidenced by recent results~\cite{venkatesh2023enhancing}. Customizing the UI to highlight key code elements can also assist in unraveling complex notebook structures. Utilizing static analysis tools available in IDEs might help with entanglement by highlighting cells potentially impacted by recent changes. This can aid in dealing with non-linear execution and identify cells that may become non-reproducible after the updates. 

Another way to enhance comprehensibility is by improving notebook search capabilities. Various studies have proposed methods for classifying notebook cells~\cite{ramasamy2023workflow, zhang2022coral, venkatesh2023enhancing}, which could be leveraged to tag cells for faster navigation or group similar cells together, provided it does not disrupt the flow of the notebook. These classifications could offer an additional layer of organization by categorically grouping cells within the notebook.

\section{Conclusion}

Integrating computational notebooks into IDEs poses new challenges and opportunities in IDE design. Whether we see notebooks as standalone environments or as integrated components of existing IDEs like PyCharm or VSCode, they present numerous unique development practices that could benefit from adding new features and tools. This design space could become even more critical with the advent of large language models (LLMs), as recent studies indicate that software development is becoming more iterative and experimental with the introduction of LLMs\cite{OpenAI_GPT4_2023, van2009reading, mcnutt2023design}.
Focusing on the support of the brainstorming and experimentation process for computational notebooks integrated into IDEs will not only improve user experience with notebooks but also help IDE designers prepare for the future of programming.

\newpage
\bibliographystyle{ACM-Reference-Format}
\bibliography{paper}

\end{document}